\documentclass[letterpaper, 10 pt, conference]{cls/ieeeconf}  

\IEEEoverridecommandlockouts                              

\title{\LARGE \bf
Experimental validation of an explicit flatness-based MPC design \\ for quadcopter position tracking}

\author{Huu-Thinh Do$^{1}$ and Ionela Prodan$^{1}$
\thanks{$^{1}${Univ. Grenoble Alpes, Grenoble INP$^\dagger$, LCIS, 26000 Valence, France}.
Email: \texttt{\scriptsize \{huu-thinh.do,ionela.prodan\}@lcis.grenoble-inp.fr}
\newline
$^\dagger$Institute of Engineering and Management Univ. Grenoble Alpes. 
}
}

\newcommand{\be}{\begin{equation}}
\newcommand{\ee}{\end{equation}}

\newcommand{\bald}{\begin{aligned}}
\newcommand{\eald}{\end{aligned}}

\newcommand{\bbm}{\begin{bmatrix}}
\newcommand{\ebm}{\end{bmatrix}}

\newcommand{\bxi}{\boldsymbol{\xi}}
\newcommand{\bu}{\boldsymbol{u}}
\newcommand{\bp}{\boldsymbol{p}}

\newcommand{\bx}{\boldsymbol{x}}
\newcommand{\bv}{\boldsymbol{v}}

\newcommand{\R}{\mathbb{R}}

\newcommand{\bzeta}{\boldsymbol{\zeta}}

\newtheorem{rem}{Remark}

\newcommand{\urlvideo}{\url{https://youtu.be/jZIKi_82xZQ}}

\newcommand{\modt}[1]{\textcolor{black}{#1}}

\usepackage{tikz}
\usepackage{physics}
\usepackage{amsmath}
\usepackage{tikz}
\usepackage{mathdots}
\usepackage{yhmath}
\usepackage{cancel}
\usepackage{color}
\usepackage{siunitx}
\usepackage{array}
\usepackage{multirow}
\usepackage{amssymb}
\usepackage{tabularx}
\usepackage{extarrows}
\usepackage{booktabs}
\usetikzlibrary{fadings}
\usetikzlibrary{patterns}
\usetikzlibrary{shadows.blur}
\usetikzlibrary{shapes}

\usepackage{pgfplots}
\pgfplotsset{width=10cm,compat=1.9}
\usepackage[hidelinks]{hyperref} 
\usepackage{mathtools} 

\usepackage{multicol}
\usepackage{cite}

\definecolor{mygray}{gray}{0.9}
\definecolor{LightRed}{HTML}{FFCECE}
\definecolor{darkgreen}{HTML}{009933}
\definecolor{brightgreen}{rgb}{0.4, 1.0, 0.0}

\usepackage{tcolorbox}
\tcbuselibrary{skins}
\begin{document}

\maketitle
\thispagestyle{empty}
\pagestyle{empty}

\begin{abstract}

\modt{
Due to the nonlinearities and operational constraints typical to quadcopter missions, Model Predictive Control (MPC) encounters the major challenge of high computational power necessary for the online implementation. This problem may prove impractical, especially for a hardware-limited or small-scale setup. By removing the need for online solvers while keeping the constraint satisfaction and optimality, Explicit MPC (ExMPC) stands out as a strong candidate for this application. Yet, the formulation was usually hindered by the two main problems:
nonlinearity and dimensionality.
In this paper, we
propose an ExMPC solution for the quadcopter position stabilization by analyzing its description (dynamics and constraints) in the flat output space.}
With the former issue, the system is exactly linearized into a concatenation of three double integrators \modt{at a price of cumbersome constraints in the new coordinates}. For the latter, with a suitable characterization of these constraints, the stabilizing ExMPC can be computed for each double integrator separately. The controller is then validated via simulations and experimental tests. The proposed scheme achieves similar performance and guarantees to the state-of-the-art solution but with notably less computational effort, \modt{allowing scalability in a centralized manner}.
\end{abstract}


\section{Introduction}

With an increasingly wide application on autonomous tasks like delivery or surveillance, it is undeniable that quadcopters have proven essential thanks to their mobility and hovering ability.
Although being investigated for decades,
 optimally governing these thrust-propelled vehicles remains an interesting topic in the research community due to their trigonometric-induced nonlinearity and challenges to rigorously provide both constraint satisfaction and stability guarantees. 
Typically, to deal with the model complexity, the systems are decoupled into two layers: attitude and position control \cite{cao2015inner}. The navigation, hence, is handled in a hierarchical scheme (a.k.a. inner–outer loop control). 

Several studies have been conducted to deal with both problems by classical and modern approaches (e.g., PI/PID \cite{lei2018robust}, sliding mode \cite{xu2006sliding} or data-driven control \cite{Torrente9361343}).
Distinguishable from the others,
Model Predictive Control (MPC) provides a framework to ensure these requirements and provide optimality thanks to its receding horizon mechanism. Indeed, the vehicle's navigation has been structurally integrated into the MPC design from various standpoints in the literature (See Table \ref{tab:MPC_drone}).
For example, with 
the standard axioms of the terminal ingredients for Nonlinear MPC (NMPC) \cite{mayne2000constrained}, the works in \cite{nguyen2020stability, nguyen2020multicopter} provide
theoretical guarantees for the stabilization of both layers. Moreover, from a geometrical viewpoint, an NMPC design was proposed in
\cite{pereira2021nonlinear} to handle obstacle avoidance and aggressive trajectory tracking. 
 A computationally efficient NMPC for the position control is proposed in \cite{gomaa2022computationally} where feasibility and stability are ensured without terminal ingredients. However, 
it is well known that 
the strategy requires high computational power and long sampling time 
when the nonlinear model is employed.

Regardless, facing the issue of nonlinearity, it is customary to seek for the notions of linearization to simplify the control design.
Indeed, as an intrinsic characteristic of the vehicle,
the properties of differentially flat systems have been exploited 
as a bridge between nonlinear system analysis and linear control theory.
On one hand, with the flat representation, a feasible trajectory can be generated via curve parameterization tools.
Then, along such an integral curve, a tangent approximation can be obtained, which is then efficiently handled by scheduling the MPC \cite{prodan2013receding}. 
While the tracking is simplified to a Quadratic Program (QP), one major theoretical drawback for this approach resides on the quantification of the approximation error and characterization of the stability for the original nonlinear system.
On the other hand, with the input-output relationship provided by the flat representation, the model can be transformed to a linear equivalent system 
\cite{levine2009analysis} in a new coordinate system, called the flat output space.
Nevertheless, 
a critical difficulty rises from the complicated input constraints during the inversion between the original and the flat output space.
Particular results on sketching theses new constraints for quadcopters' thrust, attitude and angular rate can be found in \cite{nguyen2018effective}, which map the physical limits to nonlinear sets of the flat output and up to its third order derivatives. 
In \cite{thinhECC23,mueller2013model}, linear approximation of these constraints was employed with MPC, reducing the complexity of the navigation and preserving the system's
stability analysis within a linear setting. 

\begin{table}[ht]
  \centering
  \caption{MPC-based control for quadcopter systems}
    \begin{tabular}{|p{1cm}|p{6.5cm}|}
    \hline
         & Reference  \\
    \hline
    
    & {NMPC with feedback linearization local controller, computationally
complex \cite{nguyen2019stabilizing}.}\\
\cline{2-2}       & 
{fast NMPC with stability, feasibility guaranteed, strict tuning \cite{gomaa2022computationally}.}\\

\cline{2-2}       &
{QP-MPC in the flat output space, with convoluted constraint approximation \cite{mueller2013model,thinhECC23}.}\\
\cline{2-2}       \multirow{-4}[0]{=}{Position control}  & 
\textit{Explicit} solution of B\'ezier curves, no stability analysis and constraint characterization \cite{liu2015explicit}. \\
    \hline
     &     \textit{Explicit MPC}; assumption of small roll, pitch angles \cite{Jiajin8243262}.\\
\cline{2-2}     &   
{Stability and feasibility guaranteed, conservative constraint approximation with computed-torque local controller \cite{nguyen2020multicopter}.}
\\
\cline{2-2}    \multirow{-3}[0]{=}{Attitude control}    & 
{Low-cost NMPC, suboptimal formulation, hovering point approximation for system prediction \cite{zanelli2018nonlinear}.} \\
    \hline
   &  assumption of small roll, pitch angles; soft constraints \cite{wang2021efficient} . \\
\cline{2-2}        & 
{Learning MPC with disturbance rejection, incomplete
 theoretical guarantees \cite{torrente2021data}.} \\
\cline{2-2}     \multirow{-3}[0]{=}{Full dynamics}   & 
Finite horizon LQR for the approximated dynamics, constraints neglected \cite{Cohen9143426}. \\
    \hline
    \end{tabular}%
  \label{tab:MPC_drone}%
  \vspace{-0.5cm}
\end{table}%

With the goal of pursuing the latter and advancing toward a low-cost solution while maintaining both constraint and stability guarantees, the Explicit MPC \cite{bemporad2002explicit} (ExMPC) constitutes a promising candidate. The underlying idea is to convert the nonlinearity of the quadcopter's position control into a standard QP problem via the flatness-based exact linearization, with a stability guarantee via standard synthesis in MPC \cite{mayne2000constrained} and finally solve such problem within the  multi-parametric quadratic programming framework.
The online implementation includes 
locating the current state inside the look-up table computed offline as a set of critical regions, and evaluating the associated optimal control.
Although there exists related works in the literature \cite{jiajin2017design,liu2015explicit}, the discussed methods rely on
a near hovering point assumption or lack stability quantitative arguments. 
Furthermore, to the best of the authors' knowledge, although the simulation findings in these works affirms the theory, the effectiveness of the strategy has not been experimentally validated for the model under study.
Therefore, 
exploiting the exactly linearized model in closed-loop of the quadcopter's translational dynamics, we:
\begin{itemize}
    \item propose an original synthesis procedure for implementing the ExMPC to the outer loop of the quadcopter via its exact linearization from flatness with a linear description for the convoluted constraints. 
    \item test experimentally the proposed scheme over multiple nanodrones (the experiment video is available at \urlvideo) and provide the necessary steps and code for implementation.
\end{itemize}

While the flat representation was already discussed in the literature \cite{mueller2013model,thinhECC23}, we focus on the implementation of the ExMPC
and its ancillary technical prospects. More specifically, in Section \ref{sec:rep}, the quadcopter position control problem in both the state/input state and the flat output space will be recalled. Therein, the linear approximation of the new constraint set is introduced. Section \ref{sec:formulation} presents the ExMPC formulation in the flat output space. Simulations are also provided to analyze the applicability of the strategy and compare the advantage with the implicit MPC. Section \ref{sec:validation} verified the control scheme via various experimental tests. Finally, 
Section \ref{sec:conclusion} concludes and provides future directions.

\textit{Notation:} Matrices with appropriate dimension are denoted via bold upper-case letters. Bold lower-case letters represent vectors. $\|\bx\|_{\boldsymbol{P}}\triangleq\sqrt{\bx^\top \boldsymbol{P} \bx}$ denotes the weighted norm. The letter $k$ represents the signal's value at the discrete step $kt_s$ with the sampling time $t_s$.  $\mathrm{ls}(n, a, b)$ represents a set of $n$ evenly spaced real numbers over the  interval $[a,b]$. diag($\cdot$) returns a diagonal matrix formed by its arguments. $\ominus$ denotes the Pontryagin difference. $\mathrm{conv}\{\cdot\}$ denotes the convex hull of given components.

\section{System description {in the flat output space}}
\label{sec:rep}
\subsection{Quadcopter translational dynamics}
At time step $k$, the discretized position control problem of the quadcopter can be written in the following form \cite{nguyen2020stability}:
\be 
\bxi_{i,k+1} = \boldsymbol A\bxi_{i,k}
+\boldsymbol Bh_{i}(\bu_k)
\label{eq:drone_nonlinear}
\ee 
where $\bp_k = [p_{1,k}\,\,p_{2,k}\,\,p_{3,k}]^\top\triangleq [x_k\,y_k\,z_k]^\top$ and  $x,y,z$ describe the position of the quadcopter in the inertial frame; while, for $i\in\{1,2,3\}$, $\bxi_{i,k}\triangleq \bbm p_{i,k} \\ \dot p_{i,k}
 \ebm \in\R^2$ denote the states vector collecting the position components and their derivative in each axis; 
 $\bu_k\triangleq[T_k\,\,\phi_k\,\,\theta_k]^\top\in \R^3$ collects inputs of the system including the normalized thrust, the roll and pitch angles. $\boldsymbol A \triangleq \bbm1& t_s \\ 0 &1 \ebm ,\boldsymbol B\triangleq\bbm 0.5t_s^2 \\ t_s\ebm$.
 The remaining functions are defined as:
 \be 
\begin{cases}
h_1(\bu_k)=T_k(\cos\phi_k\sin\theta_k\cos\psi   +  \sin{\phi_k}\sin{\psi}), &\\
h_2(\bu_k)=T_k(\cos\phi_k\sin\theta_k\sin\psi   -  \sin{\phi_k}\cos{\psi}) ,&\\
h_3(\bu_k)=-g+T_k\cos\phi_k\cos\theta_k,
\end{cases} 
\label{eq:func_h_psi}
 \ee 
where $g$ is the gravitational acceleration, $t_s$ is the sampling time and $\psi$ represents the measured yaw angle. Moreover, 
the input $\bu_k$ is constrained
inside a set $\mathcal{U}$ defined as:
\be 
\mathcal{U}\triangleq\{\bu_k:0\leq T_k\leq T_{max}; |\phi_k|\leq \epsilon_{max},|\theta_k|\leq \epsilon_{max}\}
\label{eq:constrU}
\ee 
with 
$T_{max}>0,\epsilon_{max}\in(0;\pi/2)$ denoting the upper bounds of the thrust ($T_k$), the roll and the pitch angles ($\phi_k,\theta_k$).

 In the literature, system \eqref{eq:drone_nonlinear} is known to be {differentially flat}, i.e., there exist a coordinate change and an endogenous dynamic feedback law that linearize the model in closed-loop \cite{levine2009analysis}. Indeed, consider the variable transformation:
 \be 
\begin{aligned}
    &T_k=\sqrt{v_{1,k}^2+v_{2,k}^2+(v_{3,k}+g)^2} \\
    &\phi_k=\arcsin{\left((v_{1,k}\sin\psi - v_{2,k}\cos\psi)/T_k\right)} \\
    &\theta_k=\arctan{\left((v_{1,k}\cos\psi + v_{2,k}\sin\psi)/(v_{3,k}+g)\right)}.
\end{aligned}
\label{eq:lin_law}
 \ee
 Then, with $v_{3,k}\geq -g$, system \eqref{eq:drone_nonlinear} yields:
 \be 
\bxi_{i,k+1} = \boldsymbol A\bxi_{i,k}
+\boldsymbol B v_{i,k}, i\in\{1,2,3\}
\label{eq:linearized_dyna}
 \ee 
 where $\bv_k=[v_{1,k}\,\,v_{2,k}\,\,v_{3,k}]^\top\in\R^3$ is the new input vector of the system. 
Note that, the dynamics \eqref{eq:linearized_dyna} in the new coordinates (called {the flat output space}) is now simply composed of three double integrators. However, as a price of the exact linearization, the constraints described as $\mathcal{U}$ given in \eqref{eq:constrU} now are complicated. Indeed, the admissible set of $\bv_k$, such that $\bu_k\in\mathcal{U}$, is $\psi$-dependent, hence practically time-varying, and non-convex \cite{thinhECC23}. 
For these impractical drawbacks, we introduce a convex subset of the feasible domain as follows. 
First, from \eqref{eq:lin_law}, one can state:
\be
\begin{aligned}
    &\sin|\phi_k|=\left|(v_{1,k}\sin\psi - v_{2,k}\cos\psi)/T_k\right|\\
    & \leq |\sqrt{(v_{1,k}^2 +  v_{2,k}^2)/(v_{1,k}^2+v_{2,k}^2+(v_{3,k}+g)^2)}|
\end{aligned}
\label{eq:phi_k}
\ee 
\be
\begin{aligned}
    &\tan|\theta_k|=\left|(v_{1,k}\cos\psi + v_{2,k}\sin\psi)/(v_{3,k}+g)\right|\\
     &\leq |\sqrt{(v_{1,k}^2 +  v_{2,k}^2)/(v_{3,k}+g)^2}|
\end{aligned}
\label{eq:theta_k}
\ee 
Then, with $|\phi_k|,|\theta_k|\leq \epsilon_{max}< \pi/2$, by bounding the right-hand side of \eqref{eq:phi_k} and \eqref{eq:theta_k} with $\sin\epsilon_{max}$ and $\tan\epsilon_{max}$, respectively, we arrive to the condition:
\be 
v_{1,k}^2+v_{2,k}^2 \leq (v_{3,k}+g)^2\tan^2\epsilon_{max}  
\label{eq:theta_phi}
\ee 
Thus, collecting \eqref{eq:theta_phi}, constraint $0\leq T_k \leq T_{max}$ and the linearizing condition in \eqref{eq:linearized_dyna}, we obtain a new constraint set:
\be 
\begin{aligned}
    \mathcal{V}_c=\big\{\bv_k\in\R^3: v_{1,k}^2+v_{2,k}^2+(v_{3,k}+g)^2\leq T_{max}^2,&\\
   v_{1,k}^2+v_{2,k}^2 \leq (v_{3,k}+g)^2\tan^2\epsilon_{max}  , v_{3,k}\geq -g&\big\}.
\end{aligned}
\label{eq:Vc}
\ee  
Consequently, the control problem has now been converted to the governing of a system of three double integrators in \eqref{eq:linearized_dyna} with their inputs $v_{i,k}$ intricately restrained in $\mathcal{V}_c$ in \eqref{eq:Vc}.

\subsection{Linear inner approximation for the input constraints}
It is noticeable that, while represented by quadratic inequalities, the set $\mathcal{V}_c$ in \eqref{eq:Vc} can be also regarded as a set bounded by two surfaces: a disk of radius $T_{max} $ centered at $[0,0,-g]^\top$, and a convex cone characterized by $g$ and $\epsilon_{max}$. 
Hence,
by parameterizing the intersection ring between the two surfaces,
the boundary point of the disk and
the vertex $[0,0,-g]^\top$, the set $\mathcal{V}_c$ can be approximated as:
\be 
\tilde{\mathcal{V}}_c = \mathrm{conv}
\left\{
\begin{aligned}
    [0,0,-g]^\top, [R^\star\cos\alpha, R^\star\sin\alpha,v_3^\star]^\top&\\
    [r\cos\alpha,r\sin\alpha,\sqrt{T_{max}^2-r^2}-g]^\top&
\end{aligned}
\right\}
\label{eq:Vctilde}
\ee 
with $\alpha\in \mathrm{ls}(\ell_1,0,2\pi),r\in \mathrm{ls}(\ell_2,0,R^\star)$
for some large integers $\ell_1,\ell_2$, $R^\star\triangleq T_{max}\sin\epsilon_{max}$, $v_3^\star\triangleq T_{max}\cos\epsilon_{max}-g$.

\begin{figure}[htbp]
    \centering
    \resizebox{0.44\textwidth}{!}{\input{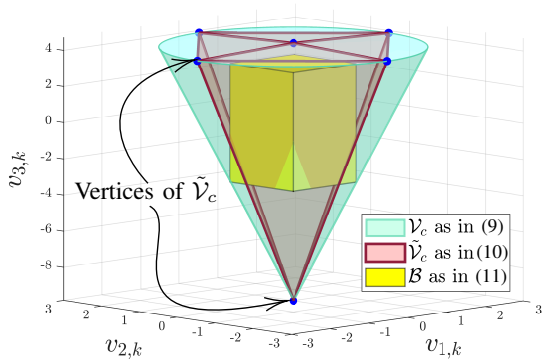}}
   \caption{The inner approximation $\tilde{\mathcal{V}}_c$ (red) of $\mathcal{V}_c$ (cyan) with $\ell_1=5,\ell_2=2$, $\epsilon_{max}=0.1745$ (rad), $g=T_{max}/1.45=9.81$ (m/s$^2$)}   
   \label{fig:VVc_para}
\end{figure}

In this fashion, the constraint set now can be regarded as linear
constraints represented by $\tilde{\mathcal{V}}_c$ in \eqref{eq:Vctilde} for some sufficiently large $\ell_1,\ell_2$.
In \cite{thinhECC23}, it was shown that the migration to the flat output space with the approximated input constraints provides a computationally attractive MPC formulation compared to nonlinear MPC (NMPC) solutions in the literature \cite{nguyen2019stabilizing,nguyen2020stability}, since the online optimization problem is now a QP for the linear dynamics and linear constraints. 
Accordingly, the stability and feasibility guarantees become more accessible
within the linear MPC design \cite{mayne2000constrained}.
With those advantages, 
in the next section, we extend our work to the implementation of the explicit MPC, with a view to opening possible integration in a low-cost embedded architecture. 



\section{ExMPC for a quadcopter position control}
\label{sec:formulation}
Previously, the problem of nonlinearity has been tackled
thanks to the mapping \eqref{eq:lin_law} and the approximation \eqref{eq:Vctilde}. With the complexity of a QP, the framework of linear ExMPC now can be applied to compute explicitly the optimal control for the six-dimensional (6D) system \eqref{eq:linearized_dyna} 
with the constraint $\bv_k\in\tilde{\mathcal{V}}_c$ as in \eqref{eq:Vctilde}. However, as notoriously known, the dimensionality burden
is no longer negligible for the dimension as large \cite{Kvasnica7810353}. 
In other words, the explicit solution is exponentially costly 
in the offline construction time, data storage capacity required and the online computation time, with respect to the dimension and prediction horizon size.

\begin{figure}[htbp]
    \centering
    \resizebox{0.475\textwidth}{!}{\input{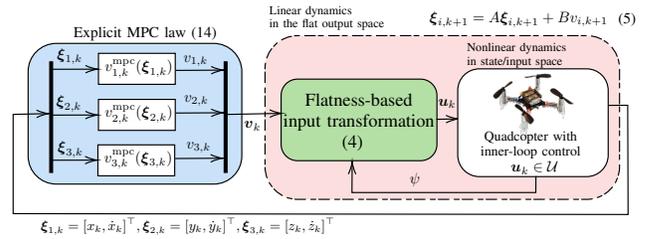}}
    \caption{Explicit MPC scheme for the quadcopter position regulation.}
    \label{fig:Scheme}
    \vspace{-0.25cm}
\end{figure}

To overcome this challenge, 
the representation of three independent double integrators in the flat output space will be exploited.
More specifically, the input constraints for $\bv_k\in \R^3$ will be approximated by a box-type constraint, giving rise to the explicit MPC formulation for each double integrator in \eqref{eq:linearized_dyna}. Comparison and simulation with the solution of the classical 6D model will be provided. A summary of the control scheme is provided in Fig. \ref{fig:Scheme}.
\subsection{Explicit MPC for the quadcopter stabilization}
Consider a box in $\R^3$ described as:
\be 
\mathcal{B}=\left\{\bv_k: |v_{i,k}|\leq \bar v_i^B, i\in \{1,2,3\}\right\} \subset {\mathcal{V}}_c \text{ as in \eqref{eq:Vc}}.
\label{eq:BoxB}
\ee 
Then, without loss of generality, we decouple the position tracking problem derived from \eqref{eq:linearized_dyna} into the three following MPC regulation problems:
\begin{subequations}
\label{eq:eMPC}
\begin{align}
&{\min } \sum_{j=0}^{N_p-1}V_s(\bxi_{i,k+j},v_{i,k+j})+V_f(\bxi_{i,k+N_p}) \label{eq:cost_J}\\
&\text{s.t:}
\begin{cases}
\bxi_{i,k+j+1}=\boldsymbol A\bxi_{i,k+j}\text{+}\boldsymbol Bv_{i,k+j},&\\
\bxi_{i,k+j} \in \mathcal{X}_i, \, j\in\{0,...,N_p-1\},&\\
|v_{i,k+j} |\leq \bar v_i^B,\bxi_{i,k+N_p} \in \mathcal{X}_{f,i}, \;i \in \{1,2,3\}&
\end{cases}
\label{eq:constr_MPC2D}
\end{align}
\label{eq:opti_1ax}
\end{subequations}
where, for each $i\in\{1,2,3\}$, $\bxi_{i,k}\in\R^2$ is the state collecting the position and the velocity of the drone projected to the three axis $\{x,y,z\}$ established as in \eqref{eq:drone_nonlinear}. $v_{i,k}\in\R$ is $i$-th component of the system's input in the flat output space described in \eqref{eq:linearized_dyna}; $\mathcal{X}_i\triangleq\{\bxi_{i}\in\R^2: |\bxi_{i}|\leq [\Bar{\mathrm{p}}_i,\Bar{\mathrm{v}}_i]^\top\}$ is the rectangular state constraint defining the workspace of the vehicle; $\mathcal{X}_{f,i}$ denotes the terminal constraint set which is associated with the terminal cost $V_f(\cdot)$ to ensure both stability and feasibility \cite{mayne2000constrained}. In details, the state cost $V_s(\cdot)$ and terminal cost $V_f(\cdot)$ are chosen as:
\be
V_s(\bxi,v)=\|\bxi\|_{\boldsymbol Q}^{2} 
+\|v\|_{\boldsymbol R}^{2} ;\;\;
V_f(\bxi)=\|\bxi\|_{\boldsymbol P}^{2}
\ee 
with $\boldsymbol R\succeq 0, \boldsymbol Q\succ 0$ and  $\boldsymbol P\succ 0$ are the user-designed weighting matrix defining the optimization problem
\eqref{eq:cost_J}. For the sake of simplicity, in \eqref{eq:cost_J}, we employ the same choice of weighting $\boldsymbol P,\boldsymbol Q,\boldsymbol R$ for all $i\in\{1,2,3\}$.

Denote $\bv_i^*(\bxi_k)\triangleq[v_{i,k}^*,...,v_{i,k+N_p}^*]^\top$ the optimizer of
\eqref{eq:opti_1ax}. Then, the control applied to system \eqref{eq:linearized_dyna}
is given as:
\be 
v_{i,k} = v^\mathrm{mpc}_{i,k}(\bxi_{i,k}) \triangleq v_{i,k}^*.
\label{eq:MPC_ex}
\ee 

Moreover, with the system's linear representation \eqref{eq:linearized_dyna}, the solution for the QP \eqref{eq:opti_1ax} can be rewritten 
in its dual form via the Karush-Kuhn-Tucker (KKT) conditions. Then, depending on 
the current state $\bxi_{i,k} $, 
a subset of constraints \eqref{eq:constr_MPC2D} 
becomes active, on which is formulated a critical region. Over this activated domain, the constrained optimum for \eqref{eq:opti_1ax} can be
explicitly parameterized by $\bxi_{i,k}$ as \cite{bemporad2002explicit}:
\be
\underset{l\in\{1,...,\bar l_i\}}{\bv_i^*(\bxi_{i,k})}
= 
    \boldsymbol F_{i,l} \bxi_{i,k}+\boldsymbol \mu_{i,l} , \text{ if }\bxi_{i,k}\in\mathcal R_{i,l};  
\label{eq:PWA_exMPC_v}
\ee 
where $\mathcal R_{i,l} = \{\bxi: \boldsymbol A_{i,l}\bxi\leq \boldsymbol b_{i,l}\}\subset \R^2$ denotes the $l$-th polyhedral critical region, and the constant parameters $\boldsymbol F_{i,l},\boldsymbol \mu_{i,l}
$ 
are the associated optimal parameters in such a region. $\bar l_i$ denotes the total number of critical regions for each double integrator in \eqref{eq:linearized_dyna}.
\begin{rem}
It is noteworthy that since $\mathcal{V}_c$ has the origin as its interior point, there always exist some subset
$\mathcal{B}$ in form of \eqref{eq:BoxB} with $\bar v_i^B$ sufficiently small. Moreover, as one candidate, the maximum volume box $\mathcal{B}$ inscribed in $\mathcal{V}_c$, can be found by inflating a zonotope via an optimization problem presented in \cite{thinhECC23}, since, $\mathcal{B}$ is, indeed, also a zonotope with the center at the origin (See Fig. \ref{fig:VVc_para}). Analytical solution for the largest $\mathcal{B}$ inscribed in $\mathcal{V}_c$ are formulated in the Appendix.
\end{rem}

\subsection{Simulation study}
\label{subsec:sim}
Previously, with the subset $\mathcal{B}$ as in \eqref{eq:BoxB}, we sidestep
the problem of dimensionality brought about by the 6D description of the model. 
To demonstrate such a computational trade-off and to better select the parameters for the proposed scheme, let us proceed by carrying out a simulation study for both approaches as the following scenarios:
\begin{itemize}
\item \textit{Scenario 1} (Sce. 1): First, we employ the proposed MPC law generated from \eqref{eq:opti_1ax}. The weighting $\boldsymbol P$ is computed from the algebraic Riccati equation. The terminal constraint set $\mathcal{X}_{f,i}$ is constructed with the classical polytopic settings for maximal positive invariant (MPI) set with a local controller $v_{i,k}^\mathrm{loc}(\bxi_{i,k})\triangleq\boldsymbol K \bxi_{i,k}$ chosen from the LQR for $(\boldsymbol A,\boldsymbol B,\boldsymbol Q,\boldsymbol R)$, ensuring the stability and feasibility of system \eqref{eq:linearized_dyna} under the control \eqref{eq:MPC_ex} \cite{mayne2000constrained}. 

    \item \textit{Scenario 2} (Sce. 2):  Herein, we concatenate the dynamics \eqref{eq:linearized_dyna} and formulate the following MPC:
    \be 
\underset{\mathclap{\bv_{k},...,\bv_{k+N_p-1}}}{\arg \min \;\;\;\;\;}  \sum_{j=0}^{N_p-1}
\|\bzeta_{k+j}\|_{\underline{\boldsymbol Q}}^2+\|\bv_{k+j}\|_{\underline{\boldsymbol R}}^2 + \|\bv_{k+N_p}\|_{\underline{\boldsymbol P}}^2
\label{eq:MPC6D}
    \ee 
 $$
\text{s.t}
\begin{cases}
{\bzeta}_{k+j+1}=\underline{\boldsymbol{A}}\bzeta_{k+j}+\underline{\boldsymbol{B}}\bv_{k+j},&\\
   \bv_{k+j}\in \tilde{\mathcal{V}}_c  \text{ as in \eqref{eq:Vctilde}},\bzeta_{k+j}\in\underline{\mathcal{X}},\bzeta_{k+N_p}\in\underline{\mathcal{X}}_f.&
\end{cases}
$$
where $\bzeta_k\triangleq [ \bxi_{1,k}^\top \,  \bxi_{2,k}^\top \,\bxi_{3,k}^\top ]^\top \in\R^6$ collects the position and velocity of the drone; $\underline{\boldsymbol{A}}=\text{diag}(\boldsymbol A,\boldsymbol A,\boldsymbol A)$; $\underline{\boldsymbol{B}}=\text{diag}(\boldsymbol B,\boldsymbol B,\boldsymbol B)$; $\underline{\mathcal{X}}=\{\bzeta_k: \bxi_{i,k}\in\mathcal{X}_i , i\in\{1,2,3\}\}$ is the same state constraints chosen as in \eqref{eq:constr_MPC2D}. Similarly, the matrices $\underline{\boldsymbol Q}\triangleq\text{diag}(\boldsymbol Q,\boldsymbol Q,\boldsymbol Q)$ and $\underline{\boldsymbol R}\triangleq\text{diag}(\boldsymbol R,\boldsymbol R,\boldsymbol R)$ are chosen from choice of weighting for the problem \eqref{eq:opti_1ax}. Lastly, the terminal ingredients $\underline{\boldsymbol P},\underline{\mathcal{X}}_f$ are computed similarly as in \textit{Sce. 1}. The explicit MPC laws is then generated in the same fashion for the optimization problem \eqref{eq:MPC6D}.
The main difference in this control synthesis is that the non-conservative approximation $\tilde{\mathcal{V}}_c$ as in \eqref{eq:Vctilde} is employed, as opposed to 
the box $\mathcal{B}$ as in \eqref{eq:BoxB} for \textit{Sce. 1}.

\item \textit{Scenario 3} (Sce. 3): In this case,
the implicit MPC law \eqref{eq:MPC6D} is implemented by 
solving the problem online. 

\end{itemize}

\begin{figure}[htp]
    \centering
    \includegraphics[width=0.45\textwidth]{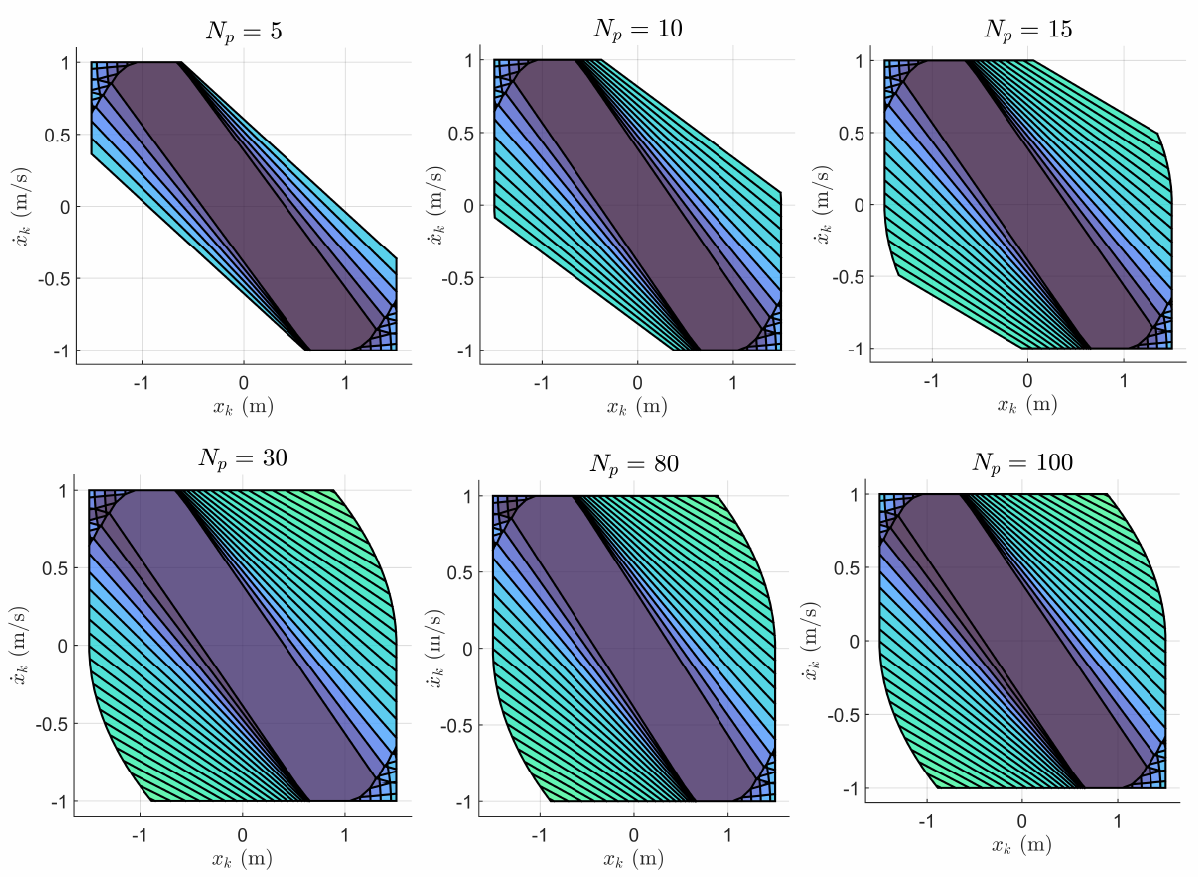}
    \caption{Explicit MPC solution with different choices of $N_p$}
    \label{fig:MultiNps}
\end{figure}

The simulations are conducted in Python 3.9.8 for the interval 60 seconds, while the critical regions are calculated with MPT3 toolbox on Matlab 2021b \cite{MPT3} and stored as a \texttt{.mat} file data. The initial position for both scenarios was chosen as: $\bxi_{1,0} = [1.25,-0.8]^\top,\bxi_{2,0} = [0,0.2]^\top,\bxi_{3,0} = [0.5,0.2]^\top$ while the tuning adopts the values: $\boldsymbol{Q} = \text{diag}(50,5),\boldsymbol{R} = 10$, resulting in the corresponding value of $\boldsymbol P=
  \bbm 524.37 & 223.75 \\
  223.75 & 225.97\ebm$. The parameter setup is given in TABLE \ref{tab:Para_sim}. 
Implementation code can be found in the following address \url{https://gitlab.com/huuthinh.do0421/explicitmpc-for-quadcopters}.

\begin{table}[htbp]
    \centering
        \caption{Parameters for simulations and experimental tests}
    \begin{tabular}{|c|c|}
    \hline
         Parameters & Values  \\ \hline
       $T_{max};\epsilon_{max}$ in \eqref{eq:constrU}  & $1.45g\approx 14.22$ m/s$^2;0.1745$ (rad)\\ \hline 
       Sampling time $t_s$ as in \eqref{eq:func_h_psi}& 100 ms\\ \hline
       $\bar v_1^B,\bar v_2^B,\bar v_3^B$ as in \eqref{eq:BoxB}& 0.8154,0.8154,3.27\\ \hline
       $\bar{\mathrm{p}}_1,\bar{\mathrm{p}}_2,\bar{\mathrm{p}}_3$ for $\mathcal{X}_i$ in \eqref{eq:constr_MPC2D}& 1.5,1.5,1.5 (m)\\ \hline
       $\bar{\mathrm{v}}_1,\bar{\mathrm{v}}_2,\bar{\mathrm{v}}_3$ for $\mathcal{X}_i$ in \eqref{eq:constr_MPC2D}& 1,1,1.5 (m/s)\\ \hline
           \end{tabular}
    \label{tab:Para_sim}
\end{table}
\begin{figure}[htbp]
    \centering
    \includegraphics[width=0.475\textwidth]{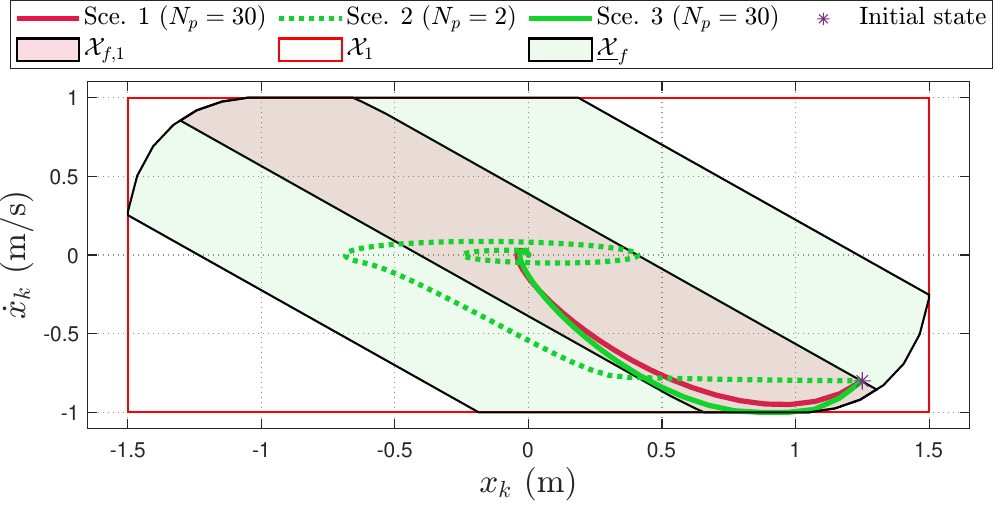}
    \caption{Comparison between the ExMPC for the proposed setup (Sce. 1), for the 6D model (Sce. 2) and the IMPC with the 6D model (Sce. 3).}
    \label{fig:ThreeSces}
    \vspace{-0.5cm}
\end{figure}
\textit{Discussion:} Fig. \ref{fig:MultiNps} presents how the critical regions expand and cover the state space with respect to different choices of the prediction horizon $N_p$. 
It is noticeable that, for a fixed choice of state constraints $\mathcal{X}_{i}$, after a certain value of $N_p$, the number of critical regions converges to a fixed value. This can also be seen in the result in Table \ref{tab:simres} for Scenario 1. After $N_p>30$, the number of regions in each of the three axes remains constant. Besides, even reaching the limit value, the computational requirement for data storage is relatively small. Meanwhile, with just
2 steps of prediction, the 6D explicit MPC solution in Scenario 2 accounts for a memory space that is 234 times larger than that of a 100-$N_p$ in Scenario 1.
Although regarded as an offline effort, the enumeration of all these regions required roughly 24 hours, in comparison with less than 3 minutes for the entire Scenario 1. Moreover, with an effortful online search (sequentially checking all the regions), the computational burden of the 6D explicit MPC is exposed via the runtime required (more than 13 times the sampling time $t_s$). This performance can be considered impractical for real-time implementation, highlighting the computational advantage of \eqref{eq:opti_1ax} over \eqref{eq:MPC6D}. \modt{
Note that during the implementation, the sequential search among the critical regions was not turned off while the system enters the invariant terminal region. Therefore, the computational strength of the ExMPC scheme necessarily
comes from the proposed decoupling settings.
}
\begin{table}[htb]
  \centering
    \caption{Simulation results and numerical specifications}
    \begin{tabular}{|p{0.72cm}|r|r|r|r|r|}
    \hline
         \multicolumn{1}{|l|}{}& 
         \multicolumn{1}{p{4em}|}{Number \newline{}of regions} & 
         \multicolumn{1}{p{0.5em}|}{Size\newline{}(MB)} & 
         \multicolumn{1}{p{0.5em}|}{$N_p$} &
         \multicolumn{1}{p{4em}|}{Avg.\newline{}CPU time}&
         \multicolumn{1}{|p{0.4cm}|}{RMS\newline{}errors}\\
          \hline \hline
    \multirow{4}[1]{=}{Sce. 1}&    (99,99,11)   &  0.02     &  5   &  0.1563 ms & 18.906 cm\\
\cline{2-6}         &    (103,103,11)  &   0.11    &   30   &   0.3385 ms&18.906  cm\\
\cline{2-6}         &   (103,103,11)   &  0.33     &    80  & 0.2864 ms &18.906  cm\\
\cline{2-6}         &   (103,103,11)  &    0.41   &     100 & 0.198 ms &18.906 cm\\
\hline\hline
    \multicolumn{1}{|l|}{Sce. 2} &       49897  &    96.0   &   2   & 1318.5 ms& 109.5 cm\\
    \hline\hline
    \multirow{4}[2]{=}{Sce. 3} &    -   &  -     &  5  &  4.72 ms & 11.46 cm\\
\cline{2-6}         &    -  &   -   &   30   &  11.63 ms&   11.319 cm\\
\cline{2-6}         &   -   &  -   &    80 & 27.05 ms & 11.319 cm\\
\cline{2-6}         &   -  &    -  &     100 & 33.08 ms &  11.319 cm\\
\hline
    \end{tabular}%
  \label{tab:simres}%
\end{table}%

Furthermore, regarding the closed-loop performance, it is certain that the Implicit MPC (IMPC) in Scenario 3 will outperform the ExMPC in Scenario 1 due to its less conservative input power (between $\mathcal{B}$ in \eqref{eq:BoxB} and $\tilde{\mathcal{V}}_c$ in \eqref{eq:Vctilde}, respectively).
This advantage is not only shown by smaller root-mean-square (RMS) errors, but also indicated via the larger terminal invariant set (See Fig. \ref{fig:ThreeSces}) for the similar tuning. 
With the interpretation as a safe region ensuring the constraint satisfaction and asymptotic stability, the terminal region $\underline{\mathcal{X}}_f$ for Scenario 3 (as well as Scenario 2) 
encloses a larger domain in the state space in comparison with Scenario 1. This indicates that, for a fixed choice of initial condition, both Scenario 2 and Scenario 3 will require a smaller size of prediction horizon to guarantee the system's origin convergence.
However, despite this advantage, the more efficient offline computation and implementation of Scenario 2 still remains open for further investigation from both combinatorial and geometric standpoints \cite{MIHAI2022308,zeilinger2011real,johansen2003approximate}.



\begin{rem}
    With the intention to adapt the stabilization problem \eqref{eq:eMPC} for trajectory tracking problem in the experiment, it is assumed that the reference trajectory satisfies:
    \begin{enumerate}
        \item    $ 
\bxi_{i,k+1}^\mathrm{ref} = \boldsymbol A\bxi_{i,k}^\mathrm{ref}
+\boldsymbol B v_{i,k}^\mathrm{ref} \text{ and } \bv_{k}^\mathrm{ref} \in \mathcal{V}_c^\mathrm{ref},
   $ where $\bxi_{i,k}^\mathrm{ref}, \bv_{k}^\mathrm{ref}$ denote the reference for the state $\bxi_{i,k}$ and the input $\bv_k$, respectively. $\mathcal{V}_c^\mathrm{ref}$ is a time-invariant set enclosing the reference signal $\bv_{k}^\mathrm{ref}$.
   \item The polyhedral set $\Delta{\mathcal{V}}_c\triangleq\tilde{\mathcal{V}}_c\ominus\mathcal{V}_c^\mathrm{ref}$ contains the origin as its interior point.
    \end{enumerate}
    
Then, with $\Delta{\bxi}_{i,k}\triangleq{\bxi}_{i,k}-{\bxi}_{i,k}^\mathrm{ref}$, the tracking error yields:
\be 
\Delta{\bxi}_{i,k+1} = \boldsymbol A\Delta{\bxi}_{i,k}
+\boldsymbol B \Delta{v}_{i,k},
\label{eq:track_err}
\ee 
with the constraint $\Delta{\bv}_k=\bv_k-\bv_k^\mathrm{ref} \in \Delta{\mathcal{V}}_c$. Consequently, by finding a box inside $\Delta{\mathcal{V}}_c$, the dynamics \eqref{eq:track_err} can be stabilized by the control synthesis for $\Delta{\bv}_k$ with the similar explicit MPC setup presented previously. Finally, the reference tracking control can be implemented as: 
\be \bv_k = \Delta{\bv}_k + \bv_k^\mathrm{ref}. \ee 
\end{rem}

\section{Experimental validation}
\label{sec:validation}
Hereinafter, we implement the control scheme described in Fig. \ref{fig:Scheme} in the Crazyflie 2.1 nano-drone framework. While the quadcopter's position and attitude are estimated via eight motion capture Qualisys cameras, the Explicit MPC
\eqref{eq:PWA_exMPC_v} is computed by carrying out a sequential search among the computed regions, respectively for all three axes (i.e, $i\in\{1,2,3\}$). The resulting control $\bv_k$ then is transformed back to the real input $\bu_k$ via \eqref{eq:lin_law} and sent to the drone 
through its Python API.
In this work, we limit our problem to the outer-loop control (position control) for the vehicle and assume that the inner loop (attitude control) is sufficiently fast and stable, or the errors between the desired angles and the real angles are negligible. Within the experimental platform, after being computed in a station computer, the desired thrust ($T_k$), the roll ($\phi_k$) and pitch angles ($\theta_k$) are stacked with the desired yaw rate $\dot\psi^\mathrm{ref}\triangleq 0$ and sent to the drone with an embedded inner-loop to follow.

\subsection{Validation setup}
For the validation, the following settings are considered.
 \begin{itemize}
     \item\textit{Test 1:} With the objective of tracking the set-point at $x_f=0.6,y_f=0.6,z_f=0.8$(m), the ExMPC computed in \eqref{eq:PWA_exMPC_v} is used. Moreover, a comparison between different choices of prediction horizon and a classical IMPC as in \eqref{eq:MPC6D} is also given. 
     \item \textit{Test 2:} Herein, four quadcopters will be controlled sequentially with the presented method to track two groups of predefined trajectories: i) \textit{Circular reference:} (Ref. 1) The four drones pursue a circular motion while maintaining a constant altitude of 1m; ii) \textit{Square formation reference:} (Ref. 2) The four drones will follow a square formation trajectories while maintaining a one-meter altitude. The trajectory generation solution was adopted from minimum length B-spline parameterization framework \cite{prodan2019necessary}.
 \end{itemize}
In the experiments, weighting matrices in Section \ref{subsec:sim} are readopted. Experiment video can be found at \urlvideo.

\begin{figure}[htbp]
    \centering
    \includegraphics[width=0.475\textwidth]{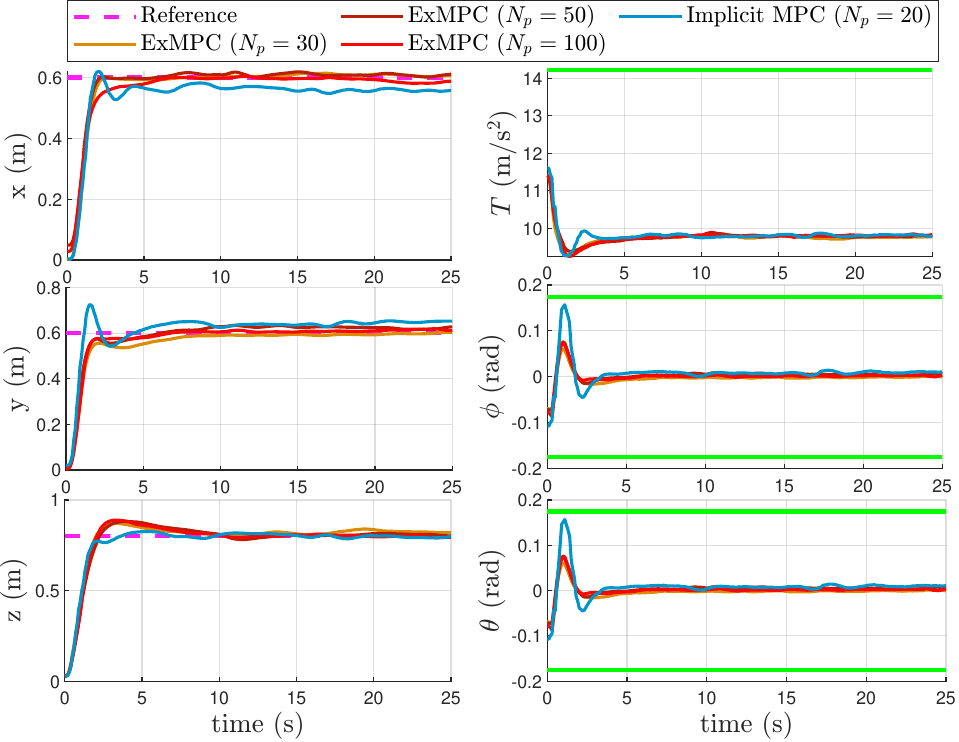}
    \caption{The quadcopter position and input $\bu_k$ (with limits shown in green solid lines) in experiment with ExMPC and IMPC.}
    \label{fig:setpts_compare}
    \vspace{-0.4cm}
\end{figure}
\begin{figure}[htbp]
    \centering    \includegraphics[width=0.45\textwidth]{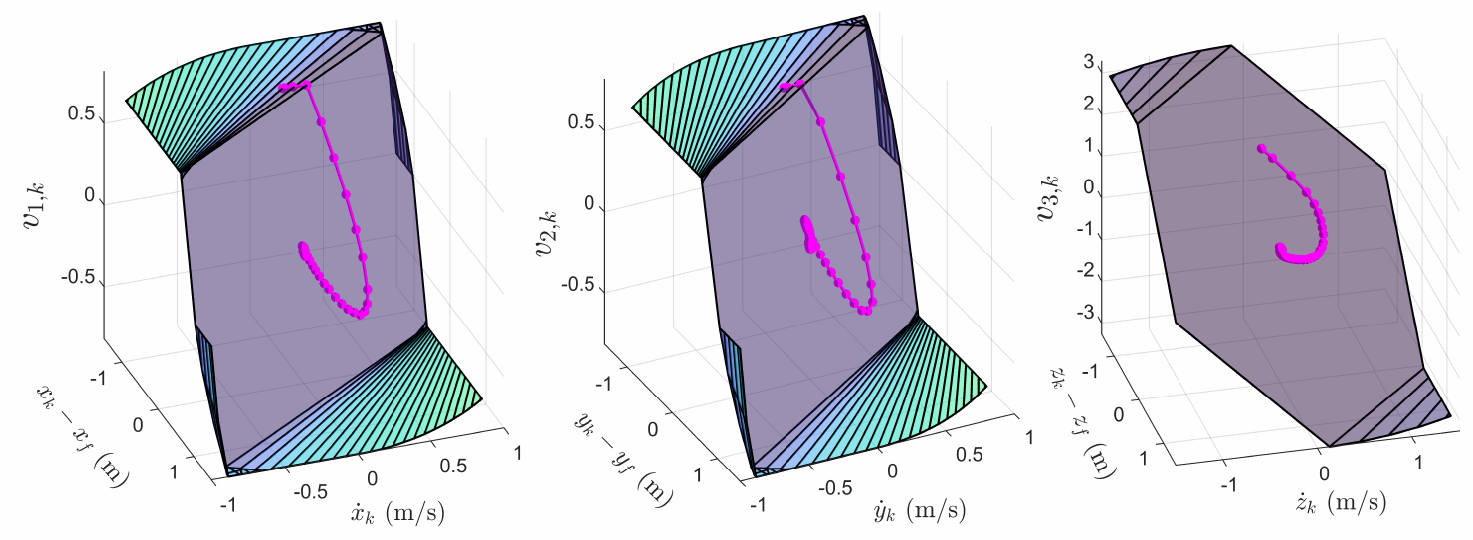}
    \caption{Experimental trajectory of $\bv_k$ in Test 1, $N_p=30$.}
    \label{fig:exp_Np30_v}
    \vspace{-0.5cm}
\end{figure}
\begin{figure}[htb]
    \centering
    \includegraphics[width=0.45\textwidth]{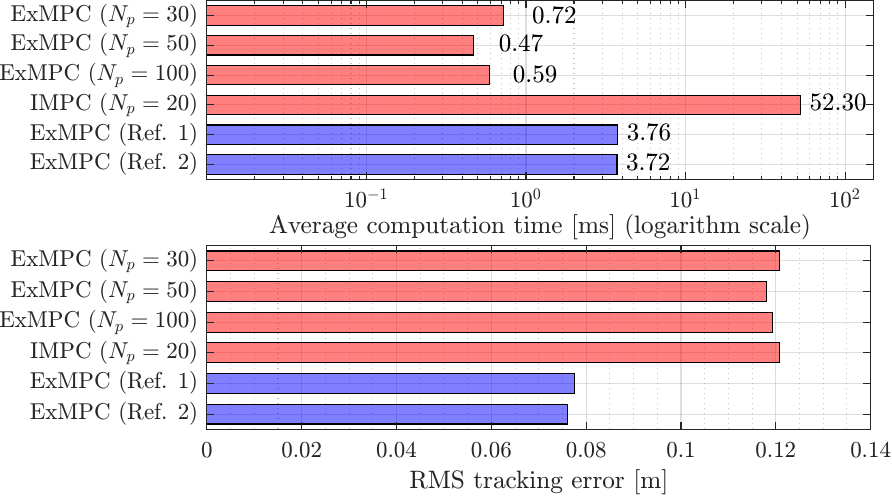}
    \caption{Computation time and RMS tracking errors with Explicit MPC and Implicit MPC in Test 1 (red) and Test 2 (blue).}
    \label{fig:Compare_OnOff}
    \vspace{-0.25cm}
\end{figure}

\begin{figure}[htbp]
    \centering
    \includegraphics[width=0.45\textwidth]{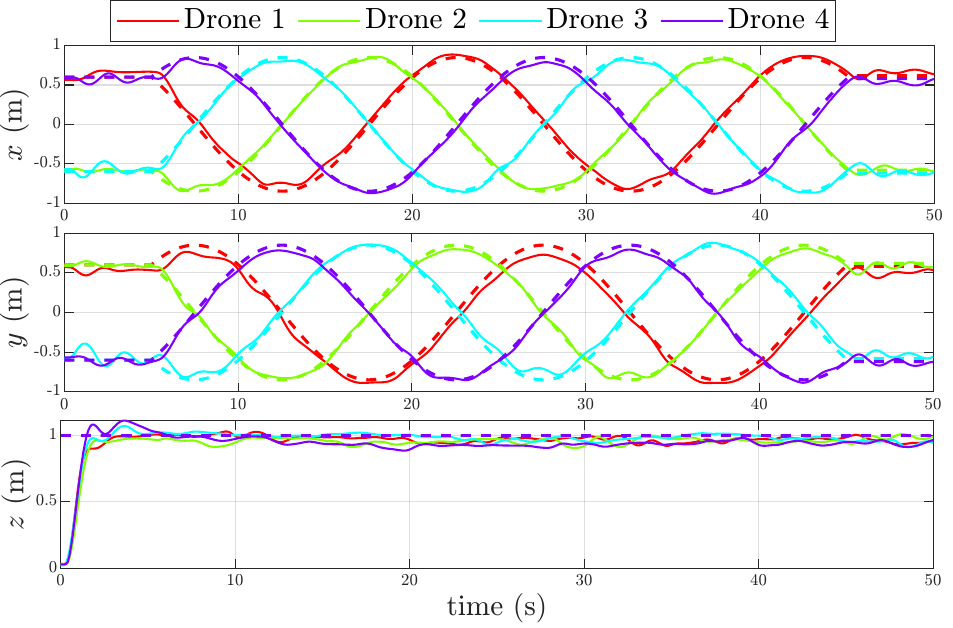}
    \caption{Multiple quadcopters' trajectory tracking with Explicit MPC (Circular reference, dashed and color-coded with the corresponding drone).}
    \label{fig:Cir_trajtrack}
    \vspace{-0.25cm}
\end{figure}
\subsection{Experimental results and discussion}
Fig. \ref{fig:setpts_compare} depicts the tested results for the Crazyflie quadcopter, where the tracking objective and input constraints are both respected.
The constraint satisfaction can also be shown by the containment of $\bv_k$ in $\mathcal{B}$ with Fig. \ref{fig:exp_Np30_v}.
The computational advantage of the proposed scheme can also be observed via the execution time for each controller in Fig. \eqref{fig:Compare_OnOff}. Therein, the IMPC demands virtually 80 times more of computation time in Test 1, although the tracking performance between the two controllers appears to be commensurate.
Even 
when there are four drones being controlled in Test 2, the computation time of the ExMPC (around 4ms) is still lower than that of the IMPC tracking with a single drone in Test 1 (52.3ms).
Similar comparisons can also be
addressed with the NMPC approaches proposed for the same guarantees in the literature \cite{nguyen2020stability,gomaa2022computationally} ($\approx$10-100ms) while the ExMPC requires under 1ms.

\begin{figure}[htbp]
    \centering
\includegraphics[width=0.47\textwidth]{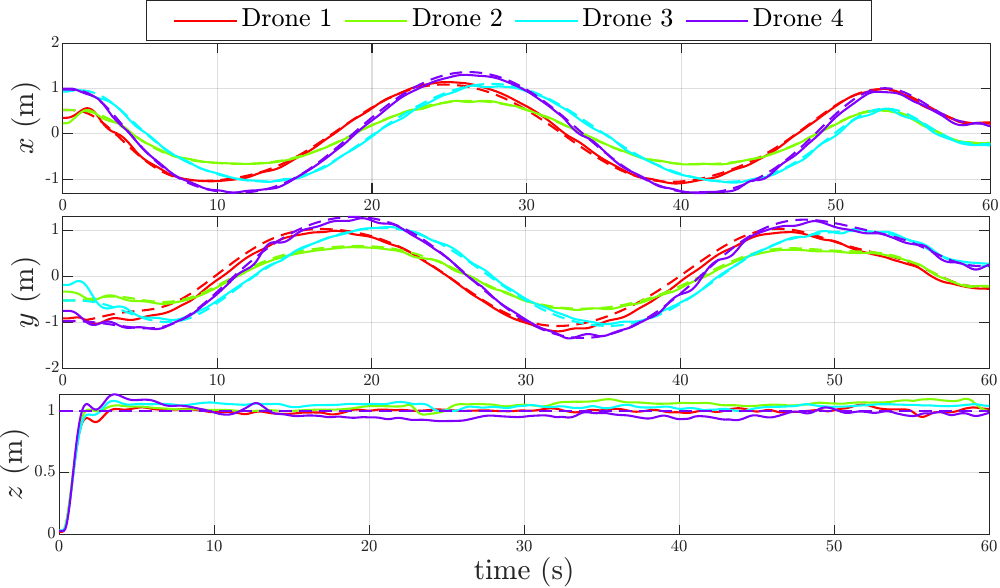}
    \caption{Quadcopters' trajectory tracking with Explicit MPC (Square formation reference, dashed and color-coded with the corresponding drone)}
    \label{fig:Sq_trajtrack}
    \vspace{-0.1cm}
\end{figure}

Finally, although there are no collision avoidance constraints imposed, the experiments show the controller's reliability to closely follow a predefined trajectory with a low computational cost, input constraint satisfaction and stability guaranteed. This satisfactory tracking opens possible implementations in hierarchical structure where collision-free formation are obtained via online curve parameterization or consensus-based guidance \cite{Luis8950150} and dynamical constraints are guaranteed in the lower level.

\section{Conclusion}
\label{sec:conclusion}
With the advantage of the linear representation in the flat output space of the quadcopter's translational dynamics,
this work proposes the constrained position tracking problem of the vehicle within the framework of the explicit MPC design. 
The successfully validated results highlight the computational advantage of the explicit solutions, while the conservativeness of the approximated input constraints can be compensated via appropriate tuning and a sufficiently large prediction horizon. Future direction concerns the implementation of more efficient point location algorithms to make use of the less conservative constraint set $\tilde{\mathcal{V}}_c$ in \eqref{eq:Vctilde}, while respecting the high sampling frequency of the vehicle.

\section*{APPENDIX}
With the zonotopic setup in \cite{thinhECC23}, the maximum volume set $\mathcal{B}$ as in \eqref{eq:BoxB} inscribed in $\mathcal{V}_c$ in \eqref{eq:Vc} can be found by solving:
\begin{subequations}
	\begin{align}
		&(\bar v_1^B,\bar v_2^B,\bar v_3^B)^*     =\arg\operatorname{max} \bar v_1^B\bar v_2^B\bar v_3^B
		,\label{eq:opt_new_drone_a}\\
	\text{ s.t: } &  \mathcal{I}(\bar v_1^B,\bar v_2^B,\bar v_3^B) \in \mathcal{V}_c \text{ as in \eqref{eq:Vc}}
\label{eq:opt_new_drone_b}
	\end{align}
	\label{eq:opt_new_drone}
\end{subequations}
where $\mathcal{I}(\bar v_1^B,\bar v_2^B,\bar v_3^B) \triangleq \{ \textstyle\sum_{i=1}^{3}0.5\gamma_i\bar v_i^B\boldsymbol{e}_i:|\gamma_i|\text{ = }1\}$ denote the finite subset of $\mathcal{B}$ which contains all of its vertices, with $\boldsymbol{e}_i$ denoting the $i$-th column of the $3\times 3$ identity matrix.
The finite condition \eqref{eq:opt_new_drone_b} implies and is implied by the containment $\mathcal{B}\subset\mathcal{V}_c$ thanks to the convexity of $\mathcal{V}_c$ while the cost function \eqref{eq:opt_new_drone_a} maximizes the volume of $\mathcal{B}$.
Moreover, with the continuity of the cost function and the boundedness of the constraints in \eqref{eq:opt_new_drone}, 
the existence of an optimal solution is certain.
By analyzing the KKT necessary conditions, the description of the set can be briefly expressed as follows:
\begin{itemize}
    \item {if $2c^\star(g+c^\star) - \tan(\epsilon_{max})^2(c^\star - g)^2<0$, then:}
\be 
    \bar v_3^B=c^\star; \bar v_1^B=\bar v_2^B= \sqrt{c^\star(c^\star+g)}
\ee 
 \item {Otherwise,}
\end{itemize}
 \be 
    \bar v_3^B=g/3; \bar v_1^B=\bar v_2^B= \tan(\epsilon_{max})\sqrt{\bar v_3^B(g-\bar v_3^B)}.
\ee 
where $c^\star \triangleq \left(-2g +\sqrt{g^2+3T_{max}^2}\right)/3$.



\end{document}